\documentclass[prl,twocolumn,showpacs,preprintnumbers,amsmath,amssymb]{revtex4}

\usepackage{graphicx}
\usepackage{color}
\usepackage{amsmath}
\usepackage{epstopdf}
\usepackage{amsbsy}
\usepackage{dcolumn}
\usepackage{bm}

\begin{document}

\title{Measuring spin correlations in optical lattices using superlattice potentials}
\author{K.\ G.\ L.\ Pedersen$^1$, B.\ M.\ Andersen$^1$, G.\ M.\ Bruun$^2$, O. F. Sylju\aa sen$^3$, and A. S. S\o rensen$^1$} 
\affiliation{
$^1$Niels Bohr Institute, University of Copenhagen, DK-2100 Copenhagen \O, Denmark\\
$^2$Department of Physics and Astronomy, Aarhus University, Ny Munkegade, DK-8000 Aarhus C, Denmark\\
$^3$Department of Physics, University of Oslo, P.~O.~Box 1048 Blindern, N-0316 Oslo, Norway}
\date{\today{}}

\begin{abstract}
We suggest two experimental methods for probing both short- and long-range spin correlations of atoms in optical lattices using superlattice potentials. The first method involves an adiabatic doubling of the periodicity of the underlying lattice to probe neighboring singlet (triplet) correlations for fermions (bosons) by the occupation of the new vibrational ground state. The second method utilizes a time-dependent superlattice potential to generate spin-dependent transport by any number of prescribed lattice sites, and probes correlations by the resulting number of doubly occupied sites.  For experimentally relevant parameters, we demonstrate how both methods yield large signatures of antiferromagnetic (AF) correlations of strongly repulsive fermionic atoms in a single shot of the experiment. Lastly, we show how this method may also be applied to probe $d$-wave pairing, a possible ground state candidate for the doped repulsive Hubbard model.

\end{abstract}
\maketitle

The study of ultracold atoms in optical lattices has produced several groundbreaking results including the observation of quantum phase transitions to a Mott state~\cite{Mott}, fermionic pairing~\cite{Chin}, 
 and fermionization of bosons in one dimension~\cite{Paredes}. 
A major, but presently unrealized goal, is to study quantum magnetism using atoms in optical lattices. For this purpose, it is important to have efficient methods to measure the atomic correlations. Possible experimental probes include Bragg scattering~\cite{Corcovilos} and quantum noise spectroscopy, where higher order correlation functions are measured by analyzing the fluctuations in time-of-flight~\cite{Altman,BSPADS,Andersen} or light polarization experiments \cite{BADS}. Time-of-flight experiments have been implemented experimentally to observe bosonic/fermionic bunching/anti-bunching and pairing \cite{TOFexps}. The experimental signature for the spin correlations with this method is, however, very small and requires averaging over many shots of the experiment.
Recent experiments have achieved single site resolution in probing optical lattice systems in two dimensions (2D)~\cite{ShersonBakr,Weitenberg}, which could be used to probe correlations. 

\begin{figure}[b]
\includegraphics[clip=true,width=1\columnwidth]{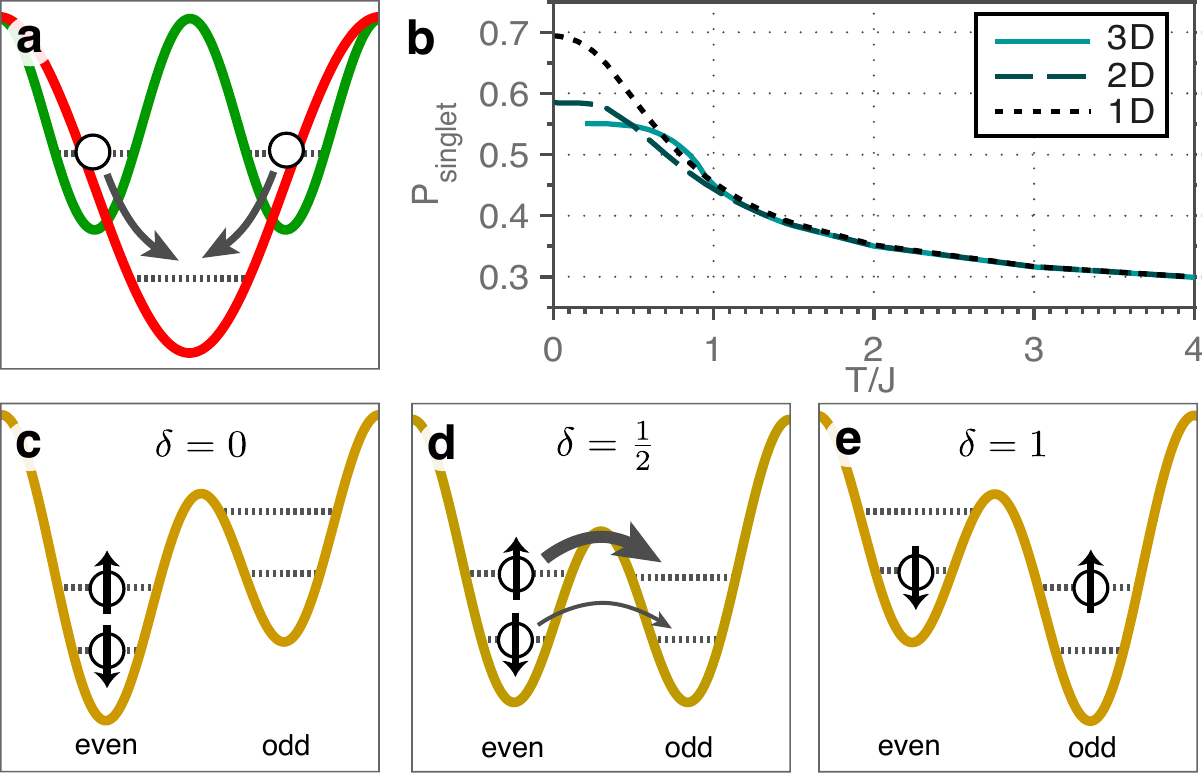}
\caption{(Color online) (a) Adiabatically doubling the lattice wavelength forces the singlet (triplet) to enter the ground state for fermions (bosons). (b) Singlet probability for nearest-neighbors vs.\ $T$ for the isotropic Heisenberg model. (c-e) A moving superlattice induces Landau-Zener transitions between
 neighboring lattice sites for spin-up atoms which have been transferred to the first excited states where the tunneling is higher than in the ground state.  Atoms on even (odd) sites move right (left) following the superlattice minima (maxima). 
}
\label{fig:idea}
\end{figure}

In this Letter, we investigate two methods for measuring correlations in optical lattices of any dimensions using superlattice potentials. Most importantly, these techniques allow for read-out of the correlations in the {\it mean value} of the relevant observable, and hence measure the correlations in a single shot of the experiment. The first method was realized in Ref. \cite{Trotzky}, and is sketched in Fig. \ref{fig:idea}(a). A superlattice is slowly turned on while the original lattice is turned off thereby merging two neighboring lattice sites. As shown below, a measurement of the population of the new vibrational ground state directly provides the nearest-neighbor singlet population. This method is ideal for measuring short-range correlations typically appearing at high temperatures $T$.  Related methods were presented in \cite{Gorelik,Jordens} allthough with much lower signals.
To detect phase transitions, we introduce a new method to probe  long-range correlations based on superlattice-induced spin-selective particle transfer by a controllable number of lattice sites, using the procedure illustrated in Fig 1(c-e). A subsequent measurement of the fraction of doubly occupied sites~\cite{Mott,Greif} reveals the long-range spin correlations. As  particular examples, we demonstrate how the method can detect AF and $d$-wave pair correlations for two-component fermions in a lattice.

We consider atoms with two internal levels $\left| \uparrow\right\rangle $ and $\left| \downarrow\right\rangle$ trapped in a cubic  lattice of period $a$. It is sufficient to consider a single dimension $x$, assuming  the motion in the other two dimensions to be  frozen by large lattice potentials. Along the $x$-direction, the atoms are trapped by a time-dependent potential given by $A(x,t) = A(t) \sin^2(\pi x/a)$\cite{Bloch}. In addition, we superimpose a superlattice potential of the form $B(x,t) = B(t) \sin^2 (\pi x/2a-\pi\delta/2)$. 
For measuring short-range correlations, two neighboring lattice sites are merged by adiabatically decreasing $A(t)$ and ramping up $B(t)$  with 
$\delta=1/2$ (see Fig. \ref{fig:idea}(a)).  If this process is performed much faster than the tunneling time, but slower than the (much faster)
oscillation period of the potential wells, one will adiabatically merge  the two wells located at $x=2n$ and $x=2n+1$, $n=0,\pm1, \ldots$. Two bosonic (fermionic) atoms in a singlet (triplet) state cannot occupy the vibrational ground state of the potential well due to the Pauli exclusion principle. Hence the singlet (triplet) state adiabatically connects to the first excited vibrational state of the two atom motional wavefunction. Thus, a measurement of the population in the vibrational ground state by  expansion imaging reveals the probability for the atoms to be in the triplet (singlet) state. 

As an application of this technique, we consider a two-component Fermi gas in an optical lattice. 
For strong repulsion and half-filling,  the gas is in the Mott phase at low $T$  and is well described by the Heisenberg model
\begin{equation}
\hat H=J\sum_{\langle l,m\rangle} {\bf \hat  s}_l \cdot{\bf \hat s}_m.
\label{Heisenberg}
\end{equation} 
Here, $\hat{\mathbf s}_l$ is the spin-$1/2$ operator for atoms at site $l$ and $\left\langle l,m \right\rangle$ denotes neighboring pairs coupled by an exchange interaction $J$. In three dimensions (3D), Eq.(\ref{Heisenberg}) exhibits a phase transition to an AF ordered state at $T=T_N\simeq 0.945J$, the observation of which is currently a main goal in the study of optical lattices. However, 
current experiments reveal only limited information on the spin state of the atoms and therefore on the temperature. 

To investigate signatures of AF correlations we consider initially the high-$T$ regime, $T\gg J$, and perform a high $T$ expansion of the density matrix 
\begin{equation}\label{highTexp}
	\hat\rho=\frac{e^{-\hat H/T}}{{\rm Tr }[e^{-\hat H/T}]}\approx\frac{1}{2^N}+\frac{J}{4T}\sum_{\langle l,m\rangle} (3\hat P^s_{lm} -\hat P^t_{lm}),
\end{equation}
where $N$ is the number of atoms, and $\hat P^s_{lm}$ ($\hat P^t_{lm}$) projects onto the singlet (triplet) state of atoms $l$ and $m$. This expression shows that interactions increase the probability of neighboring pairs to be in a singlet state by $3J/4T$. The probing method illustrated in Fig. \ref{fig:idea}(a) is therefore ideal to measure the onset of AF correlations at high $T$. 

To investigate the behavior at lower $T$, we perform full Quantum Monte Carlo (QMC) simulations using the stochastic series expansion
method \cite{SSE} with directed-loop updates \cite{SS} to calculate the correlation functions $\left\langle {\bf \hat  s}_l \cdot {\bf \hat  s}_j \right\rangle$.
This method is very efficient for Heisenberg models and gives accurate results for a wide range of $T$ for large systems. Using the spin-correlations from the QMC simulations, we show in Fig.\ \ref{fig:idea}(b) the nearest-neighbor singlet probability vs.\ $T$. Evidently adiabatically merging wells is very efficient at high $T$.
It does not, however, exhibit a clear signature for the phase transition at $T_N$, since it does not probe  the onset of long-range correlations 
 given by a non-vanishing value of  $C^z_\infty=\lim_{m\rightarrow\infty} C^z_m$ with $C^z_m= \left\langle \hat s^z_l \hat s^z_{l+m} \right\rangle$.  

This leads us to the second method which is to detect $C^z_m$ by selectively moving atoms $m$ sites either to the left or right if they are in state $\left| \uparrow \right\rangle$, whereas $\left| \downarrow \right\rangle$ atoms   remain stationary. Assuming translational invariance, the probability for two atoms to be at the same site after this displacement is given by $P_d=1/4-C^z_m$. A possible method to perform such a selective displacement based on spin selective optical lattices was proposed in Ref. \cite{BrennenJaksch}
and realized in Ref. \cite{Mandel}. This procedure, however, requires lasers to be detuned less than the fine structure splitting of the atoms. For alkali fermionic atoms used in experiments this splitting is small and may cause significant heating \cite{Blochprivate}. Our method does not suffer from this limitation,
since it is based on optical superlattices which do not need to be spin dependent and therefore can have a large detuning.

The central idea is to exploit that tunneling to neighboring sites is larger for the first excited state than for the vibrational ground state of each potential well. 
To displace only the $\left| \uparrow \right\rangle$ component, these atoms are excited to the first excited state,
 using e.g. a vibrational sideband of a Raman transition to an auxiliary state
$\left| a \right\rangle$ \cite{sideband}, while the $\left| \downarrow \right\rangle$ state remains in the  ground state. To selectively move $\left|\uparrow \right\rangle$ we then ramp up  the superlattice $B(x,t)$ with $\delta=0$ corresponding to potential minima
 at the even sites $x=2na$ of the underlying lattice. If we then let the superlattice potential  move with $\delta=vt/a$, $\left| \uparrow \right\rangle$
 atoms will be dragged along undergoing Landau-Zener transitions as illustrated in Fig.~\ref{fig:idea}(c-e): the vibrational eigen-energies at even and odd sites will anti-cross at $\delta=1/2$, where
  the system basically consists of several two-level systems. 
   Hence, the probability to cross diabatically $P$ (remain on the same site) is given by  the Landau-Zener formula, 
   $P = \exp (-2\pi g^2 / \hbar |\alpha|)$, where $g$ is half the difference between the energies at the avoided crossing, and $\alpha$ is the time-derivative of this energy difference~\cite{LZ}. Since the excited states in the neighboring  wells are stronger coupled than the ground states (have larger $g$) the superlattice velocity $v_{s}$
 can be chosen so that  the vibrational ground state remains stationary ($P\approx 1$), while  the first excited state is moved ($P \approx 0$).   Atoms in $\left|\uparrow \right\rangle$ at even (odd) sites thus move to the right (left) and 
 the  superlattice acts as a spin-dependent ``conveyor belt''. 

To demonstrate the feasibility of this proposal we  model a system of $32$ sites with lattice potential depths $A = 40 E_R$ and $B = 20 E_R$. Minimizing the sums of the errors ($1-P$ for the ground state and $P$ for the excited state) yields $v_{s} = 8.95\cdot 10^{-5} a E_R / h$.
Since  $B/A$ is rather large, there are also level crossings between the ground state and the first excited state at  different sites. To ensure that these are diabatic (non-mixing) we move the superlattice with a high velocity $v = a E_R / h$ at these crossings.
 The velocity is decreased to the optimal value only in an interval $\delta t > 0.02 a / v$ around the desired site-mixing anti-crossings. The total runtime is approximately $T \approx m\cdot 250  h/E_R= m\cdot 17$ ms ($m\cdot 8 $ ms) for the experimentally relevant case of $^{6}$Li \cite{Chin} ($^{40}$K \cite{TOFexps}), with $m$ being the number of moves we wish to perform. In order to mimic an experimentally realistic situation, we start by ramping up the superlattice potential $B$, and end the procedure by a similar ramp-down. The time-dependent Schr{\"o}dinger equation
 is solved  for each Bloch state of the superlattice unit cell by numerical integration expanding  in the 12 lowest eigenstates of the unperturbed  lattice with $B = 0$. Transforming to a suitable Wannier basis allows us to extract the transition matrices $\hat U(t) = e^{i\hat H t} = \sum_{ijnm} U_{ijnm}(t) \hat a^{\dagger}_{jn} \hat a^{\phantom{\dagger}}_{im}$, where $\hat a^\dagger_{in}$ is the creation operator for the $n$th vibrational mode Wannier function located at site $i$. The time evolution operator $\hat U(t)$ is quadratic in $\hat a$ and $\hat a^\dagger$ since we 
assume that  the interaction effects  can be tuned to zero  using a Feshbach resonance. 
We utilize Floquets' theorem for periodic Hamiltonians to extract the time-evolution.

\begin{figure}
\begin{center}
\includegraphics[width=.49\textwidth]{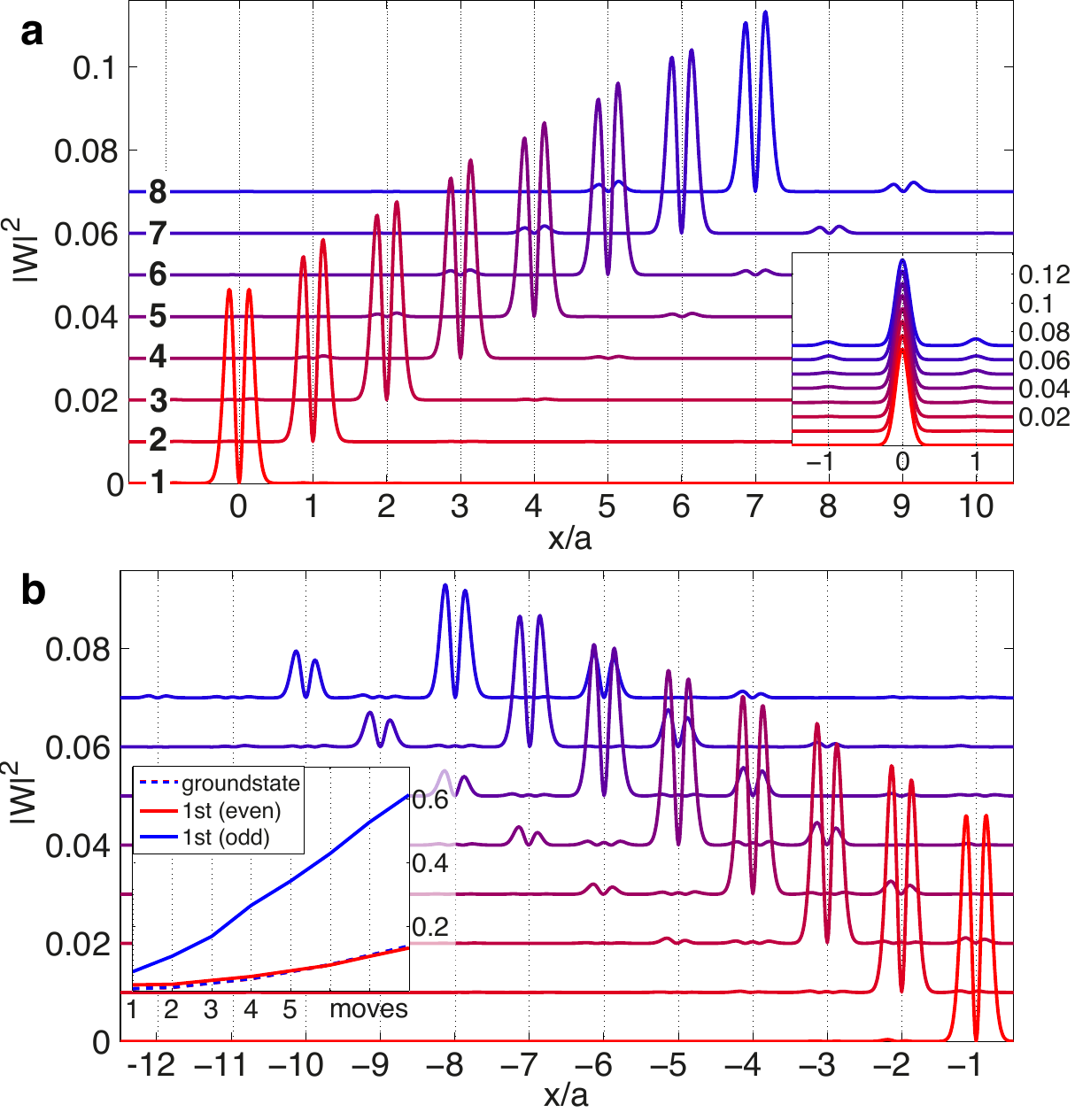}
\end{center}
\caption{(Color online) (a) Probability distribution $|W_i|^2$ of the first vibrational excitation at even sites for moves $m=1,\ldots8$. The curves are offset for clarity. The inset confirms that the vibrational ground states remains stationary. (b) Probability distribution of the first excited states on odd sites. The inset shows the error probability for the three states in (a) and (b).}
\label{FigWannier}
\end{figure}

Fig. \ref{FigWannier}(a) shows the norm square $|W_i|^2$  of the calculated Wannier functions after $m =  1,\ldots,8$ moves, for an atom starting in the first vibrational state at an even site. For this state the procedure works very well. The inset in Fig. \ref{FigWannier}(a) confirms that the ground state is immune to the moving superlattice. Figure \ref{FigWannier}(b) shows the corresponding odd site Wannier function which moves to the left.
The inset in Fig. \ref{FigWannier}(b) gives the error, i.e. the probability to be moved anything but $m$ sites from the initial position or leak to other vibrational states. Note that for the chosen parameters, the error is higher for the left moving atoms as compared to  the right moving atoms, even though the two-site Landau-Zener problem is symmetric with respect to the left and right moving atoms. This is because the left moving atoms follow the maxima of the superlattice potential $B(x,t)$ making them more susceptible for tunneling to states two sites away as seen by the growing "sidepeaks" at $m\pm2$ in Fig. \ref{FigWannier}. This resonant tunneling can be quenched by using e.g. a larger $B$ at the cost, however, of a smaller signal because of enhanced transitions to higher vibrational modes.

\begin{figure}[t]
\includegraphics[width=0.49\textwidth]{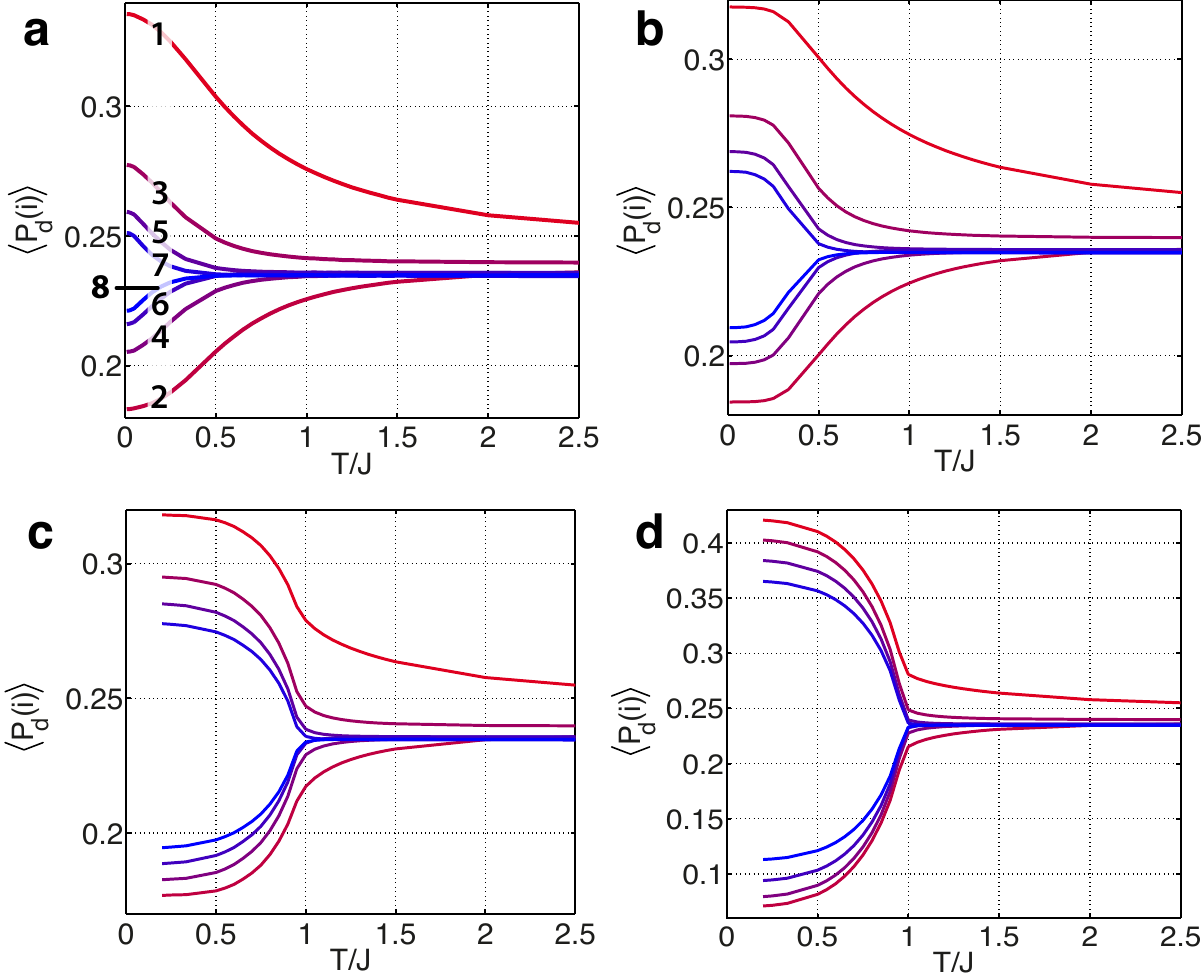}
\caption{(Color online) Average probability $\langle P_d (i)\rangle$ for a site to be doubly occupied vs. $T$ after moves $m = 1,\ldots, 8$ for the isotropic 1D (a), 2D (b), and 3D (c) Heisenberg model. (d) same as (c) with the addition of a small exchange anisotropy $H_z =0.01 J \sum_{\langle l,m\rangle} \hat  s^z_l  \hat s^z_m$.}
\label{FigPd}
\end{figure}

Assuming the (de)excitation process for $\left| \uparrow \right\rangle$ to the first vibrational state is perfect,
the probability $P_d(i)$ for site $i$ to have two atoms in the ground state $n=0$ is 
\begin{align}
& P_d(i) = \sum_{jl} \left| U_{ij11} \right|^2 \left| U_{il00} \right|^2 \left( \tfrac{1}{4}+ \tfrac{1}{2} \left\langle \hat s^z_j-  \hat s^z_l \right\rangle - \left\langle \hat s^z_j \hat s^z_l \right\rangle \right) \nonumber \\&\quad - \sum_{j\neq l} U^*_{ij11} U_{il11} U^*_{il00} U_{ij00} \left\langle \hat s^+_j \hat s^-_l \right\rangle.
\label{PdEqn}
\end{align}
Note that a measurement probing all sites leads to a $P_d$-signal proportional to the system size $N$. Due to the difference between states originating in odd and even sites, we consider the averaged $\langle P_d (i)\rangle = [P_d(i) + P_d(i+1)]/2$. 
In Fig.\ \ref{FigPd}(a-c), we show $\langle P_d (i)\rangle$ as a function of $T$ for different moves $m$ for the
 Heisenberg model in 1D, 2D and 3D. The curves are obtained from Eq.(\ref{PdEqn}) by combining the calculated transition matrices $U_{ijkl}$ for a given $m$
 with $\left\langle {\bf \hat  s}_l \cdot {\bf \hat  s}_j \right\rangle$ obtained from QMC simulations for the isotropic Heisenberg model. 
The increase in spin correlations with decreasing $T$ clearly shows up as an increase in $\langle P_d (i)\rangle$ for $m$ odd and a decrease for $m$ even. 
 The correlation decreases with increasing site separation $m$, and Fig. \ref{FigPd}(c)-(d) show how the onset of long-range order for $T<T_N$
 for a 3D system results in a dramatic change in $\langle P_d (i)\rangle$ for large $m$ as a direct consequence of the spin ordering in the AF state. Note that $\langle P_d (i)\rangle$ does not converge to the uncorrelated value $1/4$ 
 when $T/J\rightarrow\infty$ since the transfer procedure leads to atom loss into higher vibrational bands, which we do not include. As mentioned above, the pile-up of weight at the sidepeaks $m\pm2$ in Fig. \ref{FigWannier} can be minimized using other parameters, allowing a more direct comparison between $\langle P_d (i)\rangle$ and $C^z_m$. Note, however, that this contribution to the "error" does not contribute negatively to $\langle P_d (i)\rangle$ in the AF state because $m\pm2$ belong to the same sublattice as $m$.

Fig. \ref{FigPd}(d) shows $\langle P_d (i)\rangle$ vs. $T$ in 3D in the presence of a small exchange anisotropy. This case reveals the  strongest signature of the AF phase transition with a pronounced difference for large number of moves, since we force the symmetry breaking to be along the direction that we probe. By contrast, for the 1D and 2D results in Fig. \ref{FigPd}(a)-(b) the curves only ``peel off'' the $\langle P_d (i)\rangle=1/4$ line for decreasing $T$, since there is no phase transition.  Using this method, it is thus possible to  probe long-range correlations, and obtain large experimental signals in a single shot of the experiment.

Above we focused on the detection of AF correlations, but the method can be used to probe any  correlations in optical lattices.  A particularly interesting case is the possibility of $d$-wave pair correlations within the phase diagram of the repulsive Hubbard model relevant to the field of high-temperature superconductors. Such correlations are encoded in the  $d_{x^2-y^2}$-wave correlation function $C^D_{ij}=\langle \hat D_i^\dagger\hat D_j\rangle$, where 
$\hat D^\dagger_i=\hat a^\dagger_{i\downarrow}\hat a^\dagger_{i+\hat{x}\uparrow}-\hat a^\dagger_{i\downarrow}\hat a^\dagger_{i+\hat{y}\uparrow}$ with $\hat{x}$ ($\hat y$)
unit vectors along the $x$ $(y)$ direction.
To measure $C^D$, we assume that an initial operation is performed which swaps $\left| \uparrow \right\rangle$ atoms on neighboring sites 
along the $x$-axis. Next a pulse of duration $\delta t$ is applied, e.g. a photoassociation pulse, merging two atoms into a molecule if they are on the same site. This process is described by the Hamiltonian 
$
\hat H=\frac{\Omega}{2}\sum_{i} \hat b_{i}^\dagger \hat a_{i\uparrow}\hat a_{i\downarrow} +{\rm H.c.},
$
where $\Omega$ denotes the strength of the interaction and $\hat b_i$ is the bosonic annihilation operator for the molecular field. After the applied pulse, unassociated atoms are transferred back along the $x$-direction while molecules remain at rest. Neighboring $\left| \uparrow \right\rangle$ atoms  
are then swapped along the $y$-direction, and another association pulse is applied with the opposite phase $\Omega\rightarrow -\Omega$. Finally, unassociated atoms are moved back along $y$. Solving for the molecular field after the pulse we find $\hat b^{out}_{i}=\hat b^{in}_{i}-i \Omega\delta t \hat D_i/2$. Thus, the probability to find a molecule at a particular site is directly related to the probability of finding a $d$-wave pair on this and the neighboring sites. 
The total number of molecules $N_b$ is obtained  to be 
$N_b=\Omega^2\delta t^2 \sum_i C^D_{ii}/4 = \Omega^2\delta t^2 \sum_i ( 2 n_{i\uparrow} n_{i\downarrow} + 4 \left| \Delta_i \right|^2)/4$, 
where $n_{i\sigma}=\langle \hat a^\dagger_{i\sigma} \hat a_{i\sigma} \rangle$, 
and $\Delta_i=\langle \hat a_{i+\hat x\uparrow} \hat a_{i\downarrow} \rangle=-\langle \hat a_{i+\hat y\uparrow} \hat a_{i\downarrow} \rangle$ denote the 
 density and $d$-wave pairing gap, respectively. An expansion image of the molecular cloud 
should therefore reveal $d$-wave pairing as a 
 peak at zero momentum~\cite{Altman,Andersen}. The presence/absence of this peak at $T>T_c$ can also help elucidate whether the pseudo-gap phase of high-T$_c$ cuprates is caused by superconducting fluctuations or rather a hidden (non-superconducting) order. 
 
In summary, we have demonstrated how to use superlattice potentials to probe both short-range and long-range correlations in optical lattices. Specifically,  we showed how AF and $d$-wave superfluid correlations can be measured in a single shot of the experiment by this procedure. The method is, however, applicable as a probe of any correlations, and should be highly useful for future studies of quantum many-body systems.   

B.M.A. acknowledges support from The Danish Council for Independent Research $|$ Natural Sciences. O.F.S acknowledges use of NOTUR computing facilities.


\begin{thebibliography}{99}
\bibitem{Mott} M.\ Greiner \textit{et al}., Nature \textbf{415},  39 (2002); R.\ J\"ordens \textit{et al}., \textit{ibid} \textbf{455},  204 (2008);
I.\ B.\ Spielman, W.\ D.\ Phillips, and J.\ V.\ Porto, Phys.\ Rev.\ Lett. \textbf{98}, 080404 (2007);
 U.\ Schneider \textit{et al}., Science \textbf{322},  1520 (2008).
%
\bibitem{Chin} J.\ K.\ Chin \textit{et al}., Nature \textbf{443},  961 (2006).
%
\bibitem{Paredes}B.\ Paredes \textit{et al}., Nature \textbf{429},  277 (2004); T.\ Kinoshita, T.\ Wenger, and D.\ S.\ Weiss, Science \textbf{305}, 1125 (2004).
%
\bibitem{Corcovilos} T.\ A.\ Corcovilos \textit{et al}., Phys.\ Rev.\ A \textbf{81}, 013415 (2010).
%
\bibitem{Altman} E.\ Altman, E.\ Demler, and M.\ D.\ Lukin, Phys.\ Rev.\ A \textbf{70}, 013603 (2004).
%
\bibitem{BSPADS} G.\ M.\ Bruun \textit{et al}., Phys.\ Rev.\ A \textbf{80}, 033622 (2009).
%
\bibitem{Andersen} B. M. Andersen and G. M. Bruun, Phys.\ Rev.\ A \textbf{76}, 041602 (2007). 
%
\bibitem{BADS} G.\ M.\ Bruun \textit{et al}., Phys.\ Rev.\ Lett.\ \textbf{102}, 030401 (2009);
K.\ Eckert \textit{et al}., Nat.\ Phys.\ \textbf{4},  50 (2008).
%
\bibitem{TOFexps}S.\ F\"olling  \textit{et al.}, Nature \textbf{434},  481 (2005); T.\ Rom \textit{et al}., \textit{ibid} \textbf{444}, 733 (2006);
M.\ Greiner \textit{et al}., Phys.\ Rev.\ Lett. \textbf{94}, 110401 (2005).
%
\bibitem{ShersonBakr}J.\ F.\ Sherson \textit{et al}., Nature \textbf{467},  68 (2010); W.\ S.\ Bakr \textit{et al}., Science \textbf{329},  547 (2010).
%
\bibitem{Weitenberg} C.\ Weitenberg \textit{et al}., Nature \textbf{471},  319 (2011).
%
\bibitem{Trotzky} S.\ Trotzky \textit{et al}., Phys. Rev. Lett. \textbf{105}, 265303 (2010).
%
\bibitem{Gorelik} E. V. Gorelik {\it et al.}, Phys. Rev. Lett. {\bf 105}, 065301 (2010); E. V. Gorelik {\it et al.}, arXiv:1105.3356v1.
%
\bibitem{Jordens} R. J\"ordens {\it et al.}, Phys. Rev. Lett {\bf 104}, 180401 (2010).
%
\bibitem{Greif} D.\ Greif \textit{et al}., Phys. Rev. Lett. {\bf 106}, 145302 (2011).
%
\bibitem{Bloch} I. Bloch \textit{et al.}, Rev. Mod. Phys. \textbf{80}, 885 (2008).
%
\bibitem{SSE} A.~W. Sandvik and J. Kurkij\"arvi, Phys.\ Rev. B \textbf{43}, 5950 (1991).
%
\bibitem{SS} O.~F. Sylju{\aa}sen and A.~W. Sandvik, Phys.\ Rev. E \textbf{66}, 046701 (2002).
%
\bibitem{BrennenJaksch} G. K. Brennen {\it et al.}, Phys. Rev. Lett. {\bf 82}, 1060 (1999); D. Jaksch {\it et al.}, {\it ibid.} {\bf 82}, 1975 (1999).
%
\bibitem{Mandel} O. Mandel {\it et al.}, Phys. Rev. Lett. {\bf 91}, 010407 (2003).
%
\bibitem{Blochprivate} I. Bloch (private communication).
%
\bibitem{sideband} I. Bouchoule  \textit{et al.}, Phys. Rev. A \textbf{59}, R8 (1999). 
%
\bibitem{LZ} L. D. Landau, Physik. Z. Sowjet. {\bf 2}, 46 (1932); C. Zener, Proc. Royal Soc. London Series A {\bf 137}, 692 (1932).

\end{thebibliography}
\end{document}